# On the practical application of negentropic polarization of thermodynamically equilibrium photon gas

Vladimir V. Savukov

In the course of the analysis of the axiomatic principles underlying statistical physics, the assumption about the limited correctness of the postulate of equiprobability of all available microstates of a closed system was tested. Earlier, the result of simulation modeling of the most probable stationary state of a stochastic isolated system, inside which isotropic monochrome radiation spontaneously acquired anisotropic polarization, was obtained (later the presence of such an effect was confirmed by a natural experiment). This article considers an already practically significant variant of suchlike a system. The conditions for the appearance of polarization anisotropy in a thermodynamically equilibrium medium initially filled with isotropic Planck radiation are predicted. Important consistent pattern of the applied mathematical model are noted.

# РЕФЕРАТ

В ходе анализа аксиоматических принципов, лежащих в основе статистической физики, проверялось предположение об ограниченной корректности постулата равновероятности всех доступных микросостояний замкнутой системы. Ранее был получен результат имитационного моделирования наиболее вероятного стационарного состояния стохастической изолированной системы, внутри которой изотропное монохромное излучение спонтанно приобретало анизотропную поляризацию (впоследствии наличие такого эффекта было подтверждёно натурным экспериментом). В данной статье рассматривается уже практически значимый вариант подобной системы. Прогнозируются условия возникновения поляризационной анизотропии в термодинамически равновесной среде, изначально заполненной изотропным планковским излучением. Отмечены важные закономерности, характерные для применённой математической модели.

**Ключевые слова:**  дифракция, поляризация, статистическая физика, аксиоматика.

**OCIS:**  000.6590, 260.0260, 050.1940, 260.5430

**PACS:** 05.10.Ln ; 42.25.Fx ; 42.25.Ja

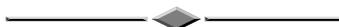

# СОДЕРЖАНИЕ

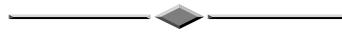

# Введение

Данная статья публикуется в рамках поискового проекта, посвящённого анализу границ применимости аксиоматических принципов статистической физики [1-3, 5, 7]. Здесь кратко излагается итог технической проработки ранее полученного результата компьютерного моделирования наиболее вероятного стационарного состояния квазизамкнутой системы, внутри которой изначально изотропное монохромное излучение в следствие дифракции на фазовой отражательной решётке спонтанно приобретало анизотропную поляризацию (позднее наличие такого эффекта было подтверждено экспериментом, выполненном на реальной установке [2]).

## Система на основе фазовой дифракционной решётки и её недостатки

Предполагалось, что вышеупомянутый эффект может быть, например, использован для решения задачи пассивной локализации объектов, находящихся в термодинамическом равновесии с окружающей средой (в скрытых системах безопасности и т. п.). Для этого данные объекты должны быть "маркированы" дифракционными решётками, поверхность которых становится видимой при наблюдении их через тепловизор с поляризационным фильтром [1]. Однако описанное техническое решение имеет ряд существенных недостатков. Главная проблема заключается в том, что анизотропия поляризационных параметров диффузного фотонного газа, возникающая после его взаимодействия с дифракционной решёткой, хорошо проявляется, если данный фотонный газ монохромный. Но состояние термодинамического равновесия характеризуется тепловым излучением с планковским распределением частот. При этом следствия от взаимодействия с решёткой различных участков спектра в очень значительной степени (примерно на 95-98%) взаимно компенсируются на уровне суммарной энергетической яркости. Ранее даже предполагалось [1], что такая компенсация может достигать 100% и в этом случае носит обязательный характер, принципиально не позволяющий тепловизорам с матрицами болометрического типа зафиксировать прогнозируемый эффект[1].

На рис. 1 приведён ряд графических изображений, иллюстрирующих данное обстоятельство. Каждый график построен в полярной системе координат так, что его центр соответствует нулевому значению угла отражения при внешнем обзоре поверхности дифракционной решётки. Величина угла отражения пропорциональна полярному радиусу, и на периферии графика значение этого угла приближается к 90°. Азимутальный угол наблюдения поверхности решётки определяется полярным углом.

Исходное световое поле представляет собой диффузное излучение с общим числом фотонов в статистическом испытании $2^{24}$ = 16777216. Индикатриса 1а описывает угловое распределение яркости монохромного (длина волны $\lambda$ = 10 мкм) диффузного светового поля, отражаемого от идеально проводящей фазовой линейной решётки (шаг $d$ = 8.200 мкм, полная глубина синусоидального профиля микрорельефа $h$ = 3.116 мкм, штрихи микрорельефа ориентированы вертикально). Этот график автоматически масштабируется так, чтобы максимальным образом выявлять все имеющиеся контрасты плотности рассеянного светового потока. На изображении индикатрисы 1а присутствуют лишь бессистемные проявления флуктуаций, в соответствии с законом Ламберта не образующие каких-либо значимых макроскопических градиентов [1].

---

[1] Данное предположение явилось причиной того, что в первой экспериментальной установке [2] объектом исследования стал предварительно стохастизированный монохромный фотонный газ, а не реальное тепловое излучение с планковским спектром.

На рис. 1б приведено изображение расчётной плотности вероятности угла поляризации α, определяемого как арктангенс отношения амплитуд взаимно ортогональных компонент электрического вектора в произвольной системе координат [6]. Этот график содержит ярко выраженные градиенты, вызываемые дифракцией монохромного излучения на отражательной решётке.

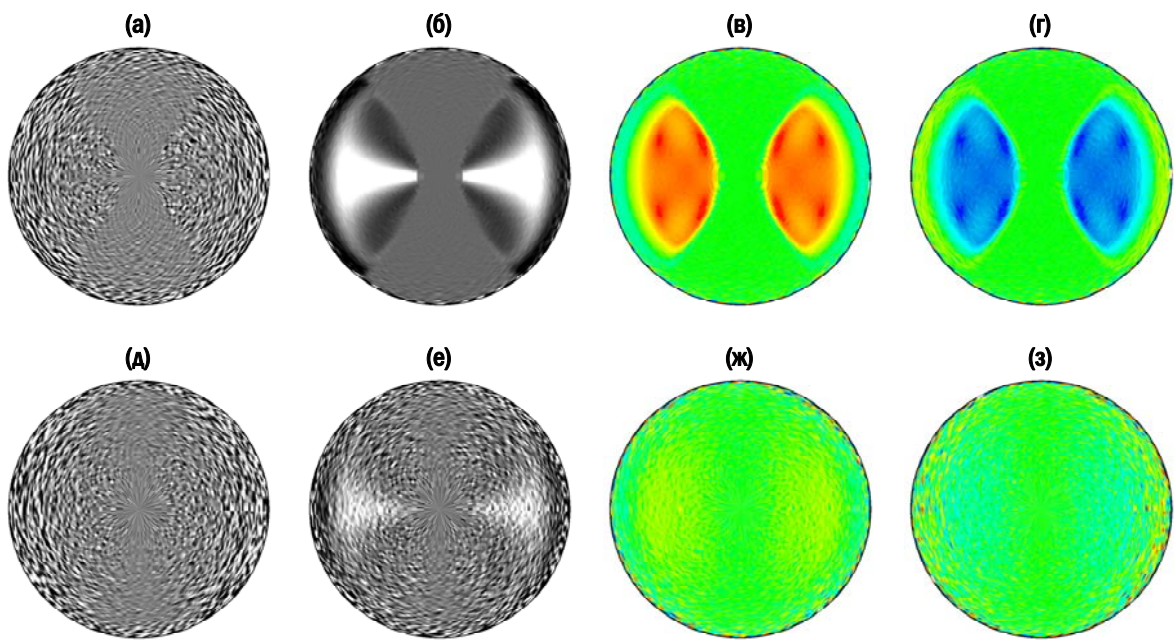

**Рис. 1.** Поляризационные характеристики изначально диффузного светового поля, рассеиваемого фазовой дифракционной решёткой с целью пассивной локации маркированного объекта:
– Верхний ряд (1а-1г) содержит информацию для монохромного светового поля;
– Нижний ряд (1д-1з) содержит информацию для светового поля с планковским спектром.

На рис. 1в и 1г представлены изображения, соответственно, *S*- и *P*-индикатрис, которые, согласно компьютерному прогнозу, можно наблюдать на экране снабжённого поляризационным фильтром тепловизора как результат дифракции диффузного излучения на решётке, но при том условии, что данное излучение является монохромным.

Для сравнения на рис. 1 приведён второй ряд графических изображений (1д, 1е, 1ж и 1з), содержащих информацию, аналогичную представленной в первом ряду (см. рис. 1а, 1б, 1в и 1г), но соответствующую случаю не монохромного, а планковского спектра изотропного фотонного газа, полностью отвечающего определению термодинамически равновесного излучения с температурой 290°К. Содержание графиков 1ж и 1з свидетельствует о том, что рассмотренная методология эксперимента практически не пригодна для работы с реальным излучением, обладающем протяжённым (планковским) спектром. Использование же на входе в тепловизор наряду с поляризационным фильтром ещё и частотного узкополосного фильтра, скорее всего, приведёт к фатальному падению величины анализируемого сигнала, исходный уровень которого уже и так невелик при типичной температуре окружающей среды (~300°К).

Из других недостатков использования дифракционных решёток для поляризации изначально термодинамически равновесного излучения следует упомянуть ещё два:

– Характер распознаваемого тепловизором сигнала сильно зависит от того, под каким ракурсом видна поверхность решётки. Есть такие сочетания углов отражения и азимутальных углов, которые образуют, своего рода, "мёртвые зоны", исключающие обнаружение поляризационных градиентов даже в среде монохромного излучения.

– Особые требования к геометрии микрорельефа дифракционной решётки могут существенно увеличивать её стоимость, поскольку наряду с шагом и глубиной этого микрорельефа регламентируется строго синусоидальная форма его профиля [2]. Кроме того, данный микрорельеф весьма уязвим перед любым внешним воздействием и может быть легко повреждён. Использование же каких-либо защитных покрытий на поверхности решётки способно существенно снизить или даже полностью исключить проявление рассматриваемых поляризационных эффектов.

### Неэргодическая система на основе плоского диэлектрического зеркала

Имеющиеся сейчас результаты имитационного моделирования наиболее вероятных стационарных состояний замкнутых физических систем дают основание предположить, что дифракционная поляризация на регулярных структурах (фазовых решётках) отнюдь не является единственно возможным механизмом, способствующим возникновению негэнтропийных процессов в таких системах. В частности, весьма перспективными могут оказаться конструкции, в которых основными оптическими элементами служат диэлектрические зеркала. В этом случае разрыв фазовых траекторий фотонов, необходимый для возникновения у системы неэргодических свойств [1], происходит в процессе взаимодействия тепловых фотонов с внешней или внутренней (собственное излучение) поверхностью зеркала.

Рассмотрим простейший вариант замкнутой системы с оптическим элементом в виде плоского диэлектрического зеркала. Анализ компьютерной модели такой системы, изначально находящейся в состоянии полного термодинамического равновесия, прогнозирует спонтанное возникновение анизотропной поляризации, заметной при наблюдении плоскости зеркала под углом отражения, равном углу Брюстера. Указанная анизотропия заключается в возникновении диспропорций между $S$- и $P$-компонентами фиксируемого излучения[1], что может быть выявлено при его надлежащей фильтрации.

Ожидаемый эффект должен быть тем сильнее, чем больше значение коэффициента преломления у материала диэлектрического зеркала. Наиболее подходящие для этой цели германий Ge и селенид цинка ZnSe имеют коэффициенты преломления, которые в пределах их окон внутреннего пропускания мало зависят от частоты излучения. Таким образом, различным частотам планковского спектра будут соответствовать примерно одни и те же величины угла Брюстера[2], т. е., в отличие от поляризации на дифракционной решётке, здесь размывание проявлений анизотропии должно быть выражено слабо.

На рис. 2 изображена схема одного из возможных вариантов экспериментальной физической установки, предназначенной для проверки описанного прогноза.

---

[1] Суммарная энергетическая яркость $S$- и $P$-компонент при этом остаётся неизменной.

[2] Угол Брюстера $\theta_{Br}$ зависит от отношения коэффициентов преломления материала диэлектрического зеркала $n_2$ и окружающей его среды $n_1$: $\theta_{Br} = \arctan(n_2/n_1)$.

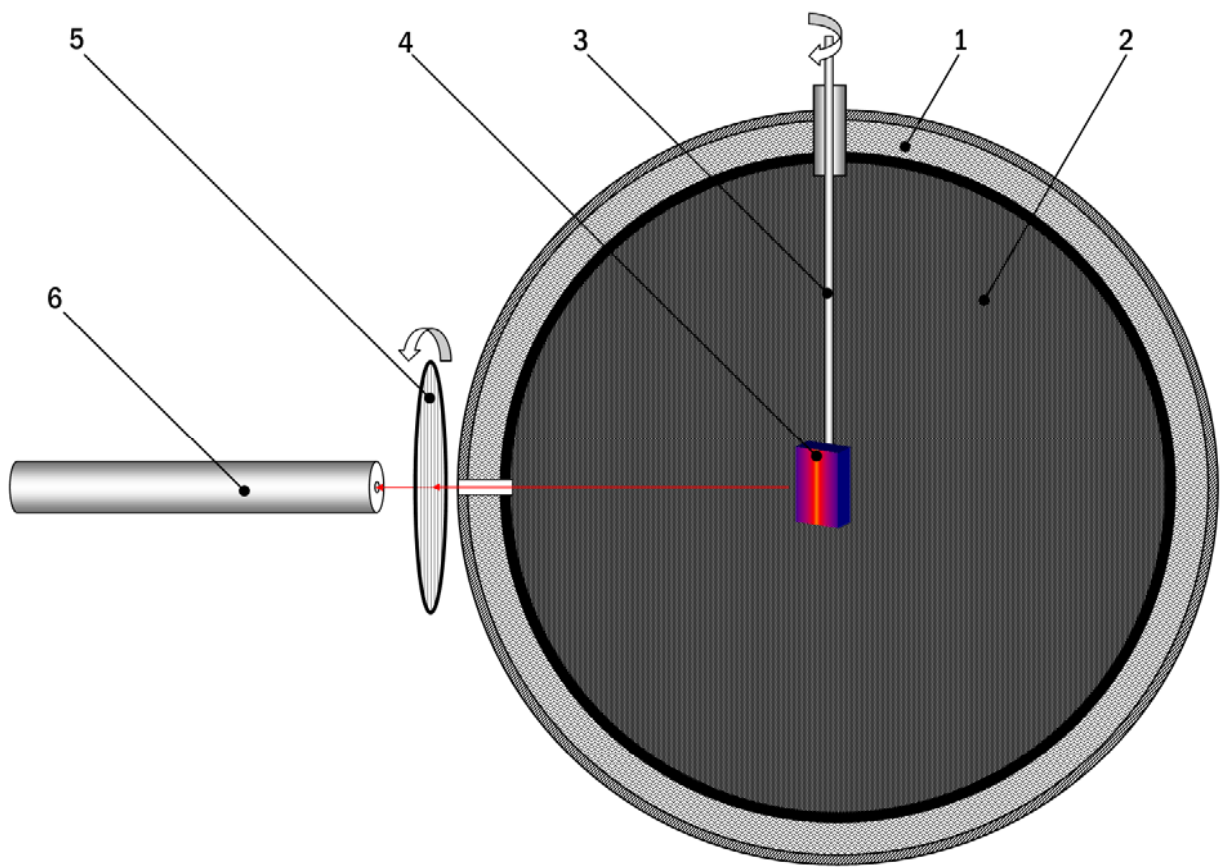

**Рис. 2.** Схема экспериментальной установки на основе плоского диэлектрического зеркала

### Перечень элементов установки:

1. Термостатированный фотометрический полый шар, ограничивающий квазизамкнутую систему. Оболочка шара, содержащая толстый слой теплоизоляции, минимизирует воздействие внешних тепловых потоков на размещённые внутри него объекты.

2. Внутренняя поверхность шара 1, покрытая поглощающим материалом со свойствами, близкими к свойствам абсолютно чёрного тела. Это, например, может быть Vantablack™ – специальная субстанция из углеродных нанотрубок, характеризуемая полным интегральным коэффициентом отражения ~ 0.045% [4].

3. Цилиндрический стержень, способный вращаться вокруг своей оси. Служит для размещения прикреплённого к нему диэлектрического зеркала 4 в центре шара 1. Изменение угла отражения, под которым оптическая ось входной апертуры пирометра 6 направлена на плоскость зеркала 4, выполняется путём поворота стержня 3.

4. Основной оптический элемент, представляющий собой плоское диэлектрическое зеркало из материала с высоким коэффициентом преломления и окном внутреннего пропускания шириной, как минимум, 8-14 мкм. Для этой цели наиболее подходит германий Ge. Важно, чтобы данный элемент не имел просветляющих покрытий!

5. Сеточный поляризатор для инфракрасной области излучения в диапазоне 8-14 мкм, желательно, на основе германия Ge или селенида цинка ZnSe. Допустим вариант и любого другого неабсорбционного поляризатора, например, проволочного типа. Данный фильтр должен иметь возможность поворота вокруг своей оптической оси. Вид компоненты излучения, пропускаемой поляризатором 5 в пирометр 6, определя-ется взаимным углом между прямой, коллинеарной направлению штрихов этого фильтра, и плоскостью диэлектрического зеркала 4.

6. Низкотемпературный (~300°K) радиационный пирометр полного излучения.

Предполагается, что данная экспериментальная установка должна зафиксировать избыток энергетической яркости $S$-компоненты или соответствующий провал яркости $P$-компоненты[1] в составе радиационного потока, регистрируемого пирометром 6 в процессе вращения зеркала 4. Эти вариации, согласно расчёту, должны, например, достигать величины ±4.0% для зеркала из селенида цинка ZnSe (угол Брюстера ≈ 67.4°) и ±5.6% – для зеркала из германия Ge (угол Брюстера ≈ 76.0°). Отклонения такого масштаба соответствуют макроскопическим градиентам температур, которые могут быть надёжно зафиксированы пирометрами даже с умеренной чувствительностью.

## Система на основе диэлектрического зеркала сферической формы

Использование диэлектрического зеркала сферической формы позволяет одновременно наблюдать на участках его поверхности с различным азимутом как избыток, так и провал яркости у поляризационных $S$- и $P$-компонент, содержащихся в составе теплового излучения, идущего со стороны указанных участков под углом Брюстера. В качестве регистрирующего прибора следует применить тепловизор, снабжённый поляризационным фильтром. Прогнозируемый эффект в данном случае должен быть одинаково хорошо заметен при любом угловом ракурсе между тепловизором и зеркалом.

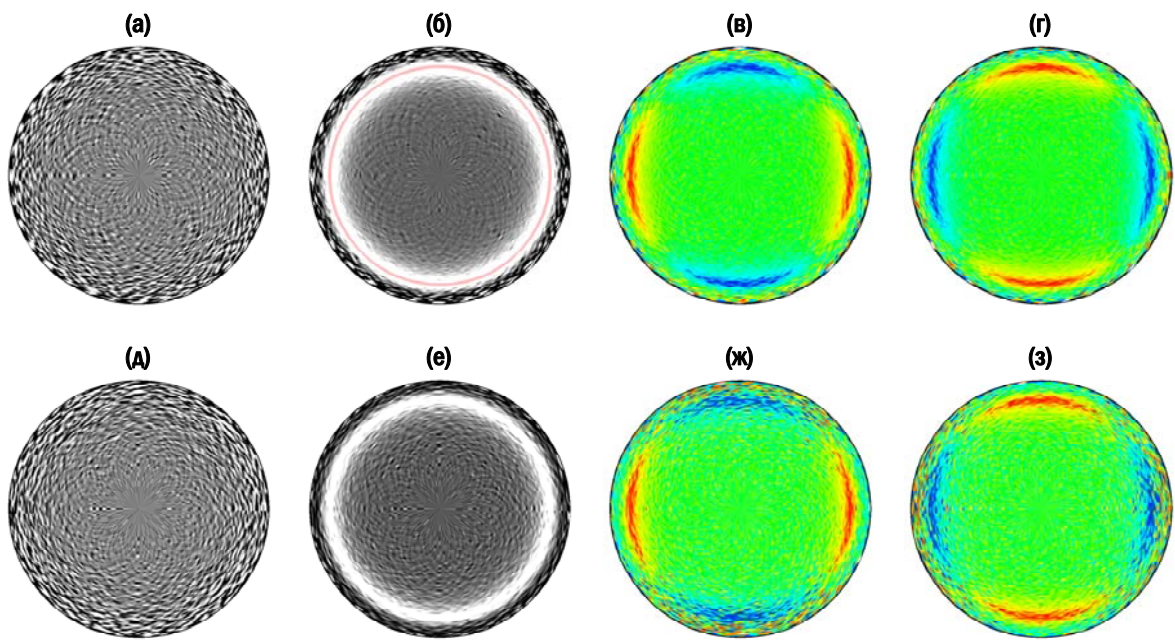

**Рис. 3.** Поляризационные характеристики, аналогичные приведённым на рис. 1, но для случая использования сферического диэлектрического зеркала вместо дифракционной решётки.

На рис. 3 приведены сравнительные результаты моделирования взаимодействия сферического германиевого зеркала как с монохромным, так и с планковским диффуз-ным излучением, причём порядок и смысловое содержание полярных графиков здесь соответствуют тем, которые были ранее представлены на рис. 1.

---

[1] Относительно их равных пропорций – в случае термодинамического равновесия.

Индикатриса 3а описывает угловое распределение яркости монохромного (длина волны λ = 10 мкм) диффузного светового поля, фиксируемого в направлении сферического зеркала. На изображении индикатрисы видны только проявления флуктуаций, не образующие каких-либо визуально наблюдаемых макроскопических градиентов.

На рис. 3б дан график расчётной плотности вероятности угла поляризации α. Это изображение содержит выраженные макроградиенты, обусловленные диспропорцией между поляризационными *S*- и *P*-компонентами в потоке излучения, идущего от зеркала под углом Брюстера[1].

На рис. 3в и 3г представлены изображения, соответственно, *S*- и *P*-индикатрис, которые, согласно компьютерному прогнозу, можно наблюдать на экране снабжённого анализатором тепловизора как результат поляризации монохромного диффузного излучения с помощью диэлектрического зеркала.

Для сравнения на рис. 3 приведён второй ряд графических изображений (3д, 3е, 3ж и 3з), содержащих информацию, аналогичную представленной в первом ряду (см. рис. 3а, 3б, 3в и 3г), но соответствующую случаю не монохромного, а планковского спектра изотропного фотонного газа, отвечающего определению равновесного излучения с температурой 290°K. Содержание графиков 3ж и 3з свидетельствует о том, что методология эксперимента с использованием диэлектрического зеркала — вполне пригодна для работы с реальным излучением, обладающем протяжённым спектром.

Эффективность проявления прогнозируемого негэнтропийного эффекта может быть проиллюстрирована на примере предполагаемых результатов применения в экспериментах диэлектрических зеркал сферической формы, выполненных из материалов с различными коэффициентами преломления "*n*". На рис. 4 представлены ожидаемые картины, которые в ранее заданных условиях (равновесное излучение, T = 290°K) можно будет наблюдать на экране тепловизора, снабжённого поляризационным фильтром.

На рис. 4а дан прогноз изображения *S*-компоненты поляризованного излучения, идущего со стороны диэлектрического зеркала, выполненного из монокристаллического германия Ge. Под этим изображением приведён график (рис. 4г) относительной яркости[2] той части излучения, которая наблюдается под углом отражения θ (вертикальной чертой отмечено значение угла Брюстера $\theta_{Br} \approx 75.98°$). Для сравнения на рис. 4б и 4д представлены аналогичные характеристики зеркала из поликристаллического селенида цинка ZnSe, а на рис. 4в и 4е – для зеркала из фтористого бария $BaF_2$. Ожидаемая вариация температур, фиксируемая тепловизором относительно исходного значения T = 290°K, составит для зеркала из германия ±5.59°K ($n \approx 4.00$), для зеркала из селенида цинка ±4.03°K ($n \approx 2.41$) и для зеркала из фтористого бария ±0.03°K ($n \approx 1.40$).

Приведённые здесь графики энергетической яркости получены на имитационной модели, использующей метод Монте-Карло с применением не псевдослучайных, а квазислучайных чисел (последовательности Соболя) [3]. Поэтому следует отметить, что изображённые на графиках 4г, 4д и 4е величины дисперсий не являются мерой достоверности соответствующих им значений математических ожиданий. Например, для германия максимум яркости $S$-компоненты характеризуется относительной величиной ≈ 0.54±0.01, т. е. реальное отклонение матожидания от термодинамически равновесного значения ½ – в четыре раза больше величины среднеквадратичного разброса. Такие пропорции означают, что только с вероятностью менее $10^{-4}$ данный максимум может быть флуктуационным проявлением равновесного состояния, т. е. достоверность существования прогнозируемых поляризационных эффектов достигает ≈ 99.994 %.

---

[1] Местоположение участка поверхности зеркала, наблюдаемого под углом Брюстера, отмечено на полярном графике красной кольцевой линией.

[2] В состоянии термодинамического равновесия относительная энергетическая яркость каждой из поляризационных компонент тождественно равна ½.

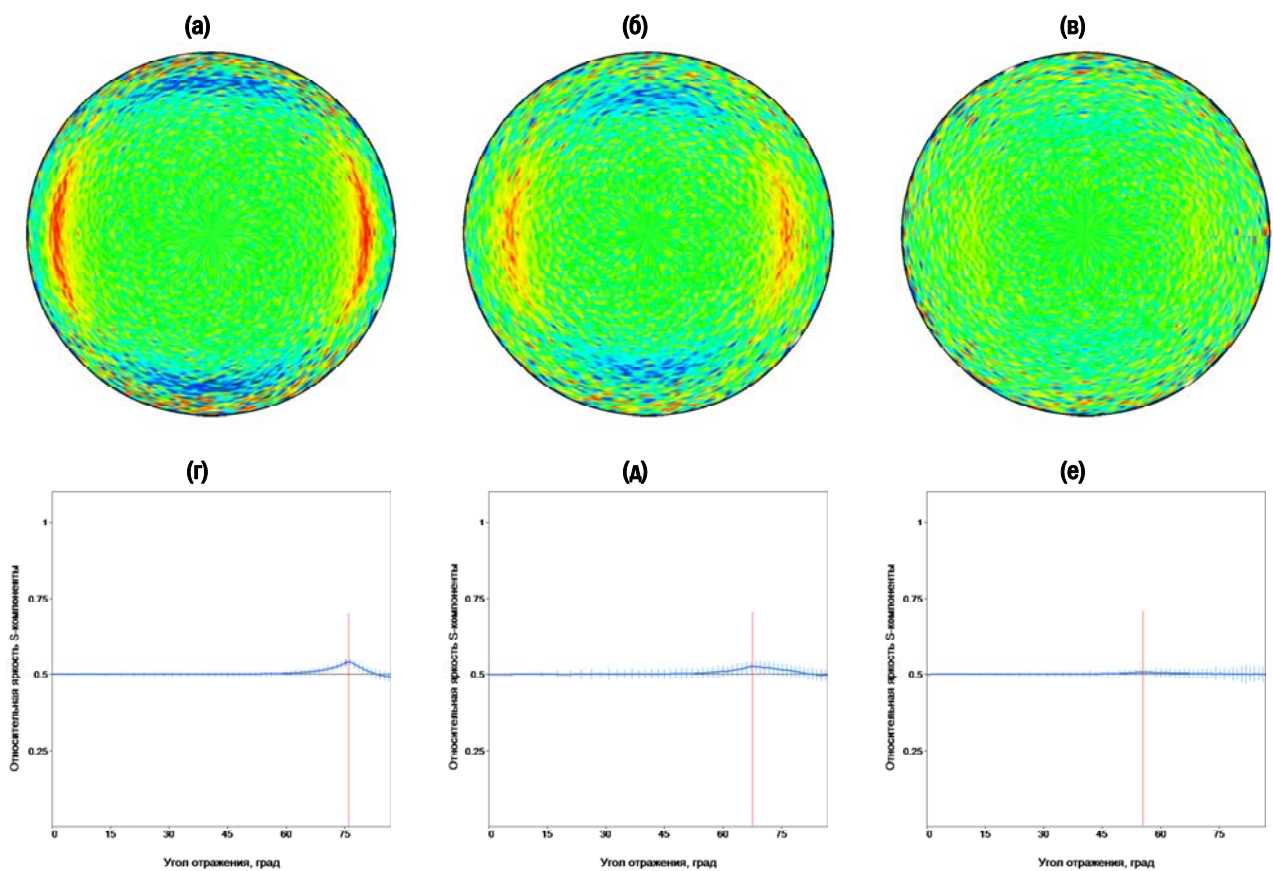

**Рис. 4.** Прогнозируемые изображения $S$-индикатрис для зеркал, выполненных из Ge ("а"), ZnSe ("б") и BaF$_2$ ("в"), а также графики радиальных сечений этих индикатрис в горизонтальной плоскости.

### Методика простейшей проверки прогнозируемого эффекта в реальных условиях

Проверка самого факта существования описанного эффекта может быть выполнено по предельно упрощённой схеме, а именно: в данном случае в качестве основного оптического элемента допустимо вместо заготовки[1] не очень распространённой сферической формы[2] использовать образец так называемого "окна", т. е. плоскую пластину из соответствующего материала (Ge или ZnSe), не имеющую каких-либо покрытий.

---

[1] До нанесения просветляющих, защитных и иных видов покрытий.

[2] Или выглядящей с определённого ракурса как сферическая (например, полусфера).

Установим данное "окно" в некотором пространстве с однородным полем температуры. Затем с помощью тепловизора и поляризационного фильтра зафиксируем яркость *S*- и *P*-компонент излучения, идущего со стороны "окна" под углом Брюстера $\theta_{Br}$. Для поляризационных компонент, наблюдаемых в окрестностях этого угла ($\approx \theta_{Br} \pm 5°$), максимальная разница в энергетической яркости ожидается равной разнице температур ±4.0°К – для "окна" из ZnSe (см. рис. 4г) и ±5.6°К – для "окна" из Ge (см. рис. 4д).

Необходимость в работе с тепловизором искать ракурс, при котором его FOV (поле зрения – Field of view) будет содержать в себе угол Брюстера (желательно, в центре диапазона FOV), немного увеличит трудоёмкость данного эксперимента. Снимки же, полученные с помощью плоского "окна", могут выглядеть не так эффектно, как в случае со сферическим зеркалом. Однако, это не уменьшает убедительности получаемых результатов, а умеренность требований к уровню используемых технических средств позволяет воспроизвести такую проверку и при наличии ограниченных возможностей.

### Обнаруженные закономерности параметров фазового пространства

Все проявления анизотропии параметров фотонного газа, ранее обнаруженные при анализе наиболее вероятных стационарных состояний замкнутых систем, обусловлены исключительно только вариациями угла поляризации α в "упруго" формируемом световом поле[1] (см., например, рис. 1б и 3б). Однако, это не означает, что изменения угла α могут носить произвольный характер. Результаты моделирования стационарных состояний замкнутых систем[2], получаемые после завершения всех переходных процессов, непременно характеризуются следующим обязательным свойством:

$$\int_0^{\pi/2} \cos(\alpha)^2 g(\alpha) d\alpha \equiv \frac{1}{2} \tag{0.1}$$

где $g(\alpha)$ - это плотность вероятности угла поляризации, т. е. нормированная к единице неотрицательная функция; для состояния термодинамического равновесия она имеет единственно возможный (в соответствии с аксиоматикой статистической физики) вид:

$$g(\alpha) = \sin(2\alpha) \geq 0, \quad \int_0^{\pi/2} g(\alpha) d\alpha = 1, \quad \forall \alpha \in [0, \pi/2] \tag{0.2}$$

Разумеется, при такой дефиниции плотности вероятности невозможно появление каких-либо градиентов у количественного соотношения *S*- и *P*-компонент в системе. Попробуем теперь ввести модифицированный вариант плотности вероятности $G(\alpha)$, который бы отвечал тождеству (0.1) и нормировке (0.2), но при этом позволял варьировать угол поляризации α максимально свободным образом:

$$G(\alpha) = g(\alpha) + R(\alpha) \geq 0, \quad \forall \alpha \in [0, \pi/2] \tag{0.3}$$

---

[1] Здесь имеется в виду взаимодействие светового поля с оптическими элементами системы, осуществляемое без необратимых потерь на поглощение (их учёт потребовал бы использования аксиоматики неравновесных процессов – гипотезы молекулярного хаоса [7]). При этом никак не нарушаются законы Кирхгоффа (баланс испускаемого и поглощаемого излучения), Ламберта (сочетание угловых распределений испускаемого и отражаемого излучений) и Планка (частотный спектр равновесного излучения).

[2] Условие (0.1) не выполняется, если система не замкнута или крайне нестационарна.

Разложим в ряд Фурье функцию $R(\alpha)$, добавленную в состав $G(\alpha)$:

$$R(\alpha) = \frac{a_0}{2} + \sum_{k=1}^{\infty}\left(b_k \sin(k\alpha) + a_m \cos(m\alpha)\right) \qquad (0.4)$$

Тогда ограничения на гармоники, разрешённые для функции $R(\alpha)$, выглядят так:

$$\int_0^{\pi/2} G(\alpha)\,d\alpha = \int_0^{\pi/2} \sin(2\alpha) + R(\alpha)\,d\alpha \equiv 1 \qquad (0.5)$$

$$\int_0^{\pi/2} \cos^2(\alpha) R(\alpha)\,d\alpha \equiv 0, \quad \forall a_m \qquad (0.6)$$

Окончательный вид функции $G(\alpha)$ с учётом условий (0.5) и (0.6):

$$G(\alpha) = \sin(2\alpha) + \sum_{m=4,6\ldots}^{\infty} a_m \cos(m\alpha) \geq 0, \quad \forall \alpha \in [0,\pi/2) \qquad (0.7)$$

На рис. 5 приведены графики для четырёх случаев распределения плотности вероятности угла поляризации $G(\alpha)$ в его области допустимых значений (ОДЗ). Кружками обозначены данные имитационных компьютерных моделей. Красными линиями выполнена аппроксимация этих данных с помощью функции $G(\alpha)$, которая содержит (0.7) лишь около одной четверти всех без исключения гармоник ряда Фурье (0.4). Тем не менее, разрешённых гармоник оказывается вполне достаточно для того, чтобы точно соответствовать моделируемым параметрам, учитывая даже проявления флуктуаций. Такой результат однозначно доказывает принадлежность рассчитываемых состояний к множеству, соответствующему условию (0.1).

Полученные данные иллюстрируют ранее сделанные выводы о различной эффективности отражающих фазовых дифракционных решёток и диэлектрических зеркал для поляризации диффузного фотонного газа с планковским спектром. На рис. 5а и 5в видно, насколько существенно отличаются плотности вероятностей $G(\alpha)$, формируемые основными оптическими элементами, от аналогичных исходных функций $g(\alpha)$ при монохромном излучении (показаны синими линиями). Для излучения с планковским спектром такие отличия плотности вероятностей большей частью сохраняются лишь при использовании диэлектрических зеркал (рис. 5г), но не фазовых решёток (рис. 5б).

Надо особо отметить значимость свойства (0.1) для замкнутых систем с той точки зрения, что, несмотря на возможность существенной анизотропии угла поляризации в геометрическом пространстве (0.7), эта анизотропия сама по себе принципиально неспособна привести к появлению градиентов температур в замкнутой системе. Хотя, например, для осмия Os разница в коэффициентах поглощения $S$- и $P$-компонент термодинамически равновесного излучения источника типа "A" (T = 2856°K) может достигать почти двукратной величины, никакая анизотропная поляризация, если она проявляется в сочетании с условием (0.1), не способна изменить коэффициент поглощения $\mu$. Это происходит потому, что зависимость данного коэффициента от угла поляризации выражается линейной параметрической функцией, где в качестве одной из переменных выступает коэффициент поглощения $\mu$, а в качестве другой — квадрат косинуса угла $\alpha$:

$$\mu(\alpha) = \mu_1 \eta(\alpha) + \mu_0, \quad \eta(\alpha) = \cos^2(\alpha) \qquad (0.8)$$

$\mu_0$ и $\mu_1$ - константы, определяющие свойства материала, поглощающего излучение.

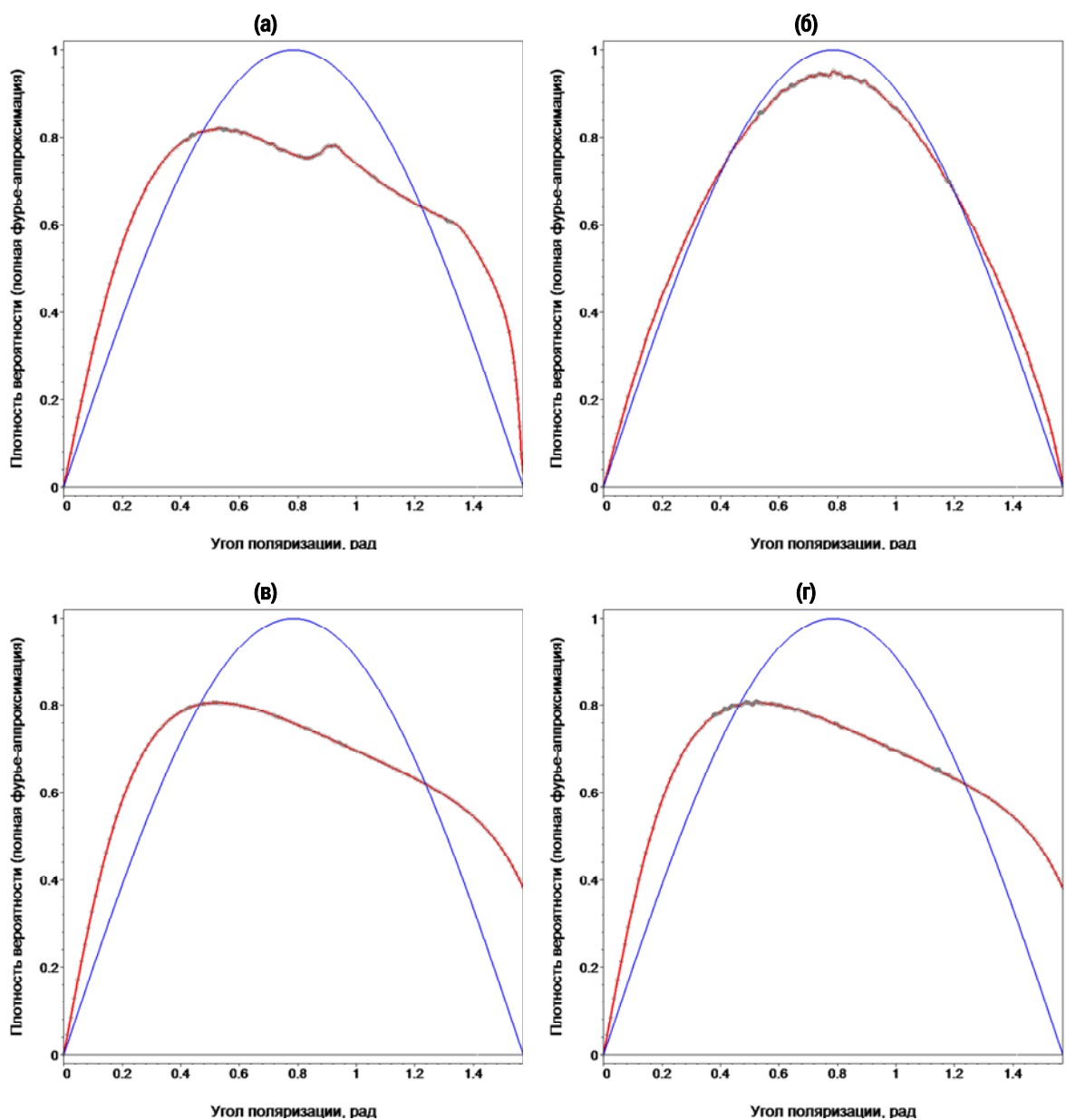

**Рис. 5.** Графики плотности вероятности угла поляризации $G(\alpha)$ в его области допустимых значений:
  а). $G(\alpha)$ - монохромный фотонный газ после рассеяния на отражающей фазовой решётке
  б). $G(\alpha)$ - планковский фотонный газ после рассеяния на отражающей фазовой решётке
  в). $G(\alpha)$ - монохромный фотонный газ после взаимодействия с диэлектрическим зеркалом
  г). $G(\alpha)$ - планковский фотонный газ после взаимодействия с диэлектрическим зеркалом

Таким образом, если в фазовом пространстве замкнутой системы реализуется условие (0.1), то величина коэффициента поглощения µ начинает совпадать со значением, характерным для изотропного излучения, не имеющего поляризационных градиентов. Это исключает возникновение разницы температур в такой системе, если только для получения температурного перепада не будет задействован какой-либо иной механизм.

Далее в Приложении в качестве гипотезы кратко рассмотрен один из таких механизмов, в основе работы которого лежит реализация условия Брэгга-Вульфа в фотонном кристалле (решётка Брэгга) при наличии слабого внутриобъёмного поглощения.

## Выводы

В процессе ранее выполненной работы был получен следующий результат: теоретически обосновано [1, 7] и практически подтверждено [2] существование неэргодических замкнутых физических систем, наиболее вероятные стационарные макросостояния которых зависят от их конфигурации. Эта зависимость выражается в разнообразии вариантов поляризационных характеристик теплового излучения, заполняющего данные системы в указанных макросостояниях.

Перспективы дальнейших исследований по данной тематике делятся на две части:

– Углублённая техническая проработка задачи пассивной локализации объектов, находящихся в термодинамическом равновесии с окружающей средой (для скрытых систем безопасности и т. п.). Такие объекты должны быть снабжены доступными для внешнего наблюдения сферическими оптическими элементами, выполненными из диэлектрика, имеющего высокий коэффициент преломления и прозрачного для теплового излучения (например, из германия). Поляризационная анизотропия, создаваемая этими элементами в изначально равновесном излучении, может быть надёжно зафиксирована с помощью тепловизора, снабжённого анализатором.

– Развитие предположения о том, что термодинамическое равновесие в общем случае не является единственно возможным наиболее вероятным макросостоянием неэргодических систем. Сказанное ставит под сомнение всеобъемлющий характер главного аксиоматического принципа статистической физики о равновероятности всех доступных микросостояний в замкнутой системе. Из этого следует, что не исключена ревизия границ корректного применения H-теоремы Больцмана, представляющей собой статистический аналог Второго закона термодинамики. О возможности данной ревизии, в частности, говорит обнаружение таких конфигураций вышеуказанных систем, в наиболее вероятных состояниях которых прогнозируется наличие градиентов температур. Разумеется, оценка подобных перспектив должна быть крайне сдержанной. Однако, до сих пор тщательная верификация исследуемой имитационной модели не выявила каких-либо принципиальных ошибок, что нашло своё подтверждение и в результатах реальных физических экспериментов. Поэтому продолжение поисковых работ в данном направлении следует считать вполне оправданным.

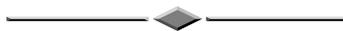

## ПРИЛОЖЕНИЕ: Брэгговская решётка в качестве основного оптического элемента

На рис. 6а приведена схема экспериментальной физической установки с брэгговской дифракционной решёткой в качестве основного оптического элемента. Показаны синусоидальный и квазипрямоугольный варианты профиля её микрорельефа (рис. 6б и 6в).

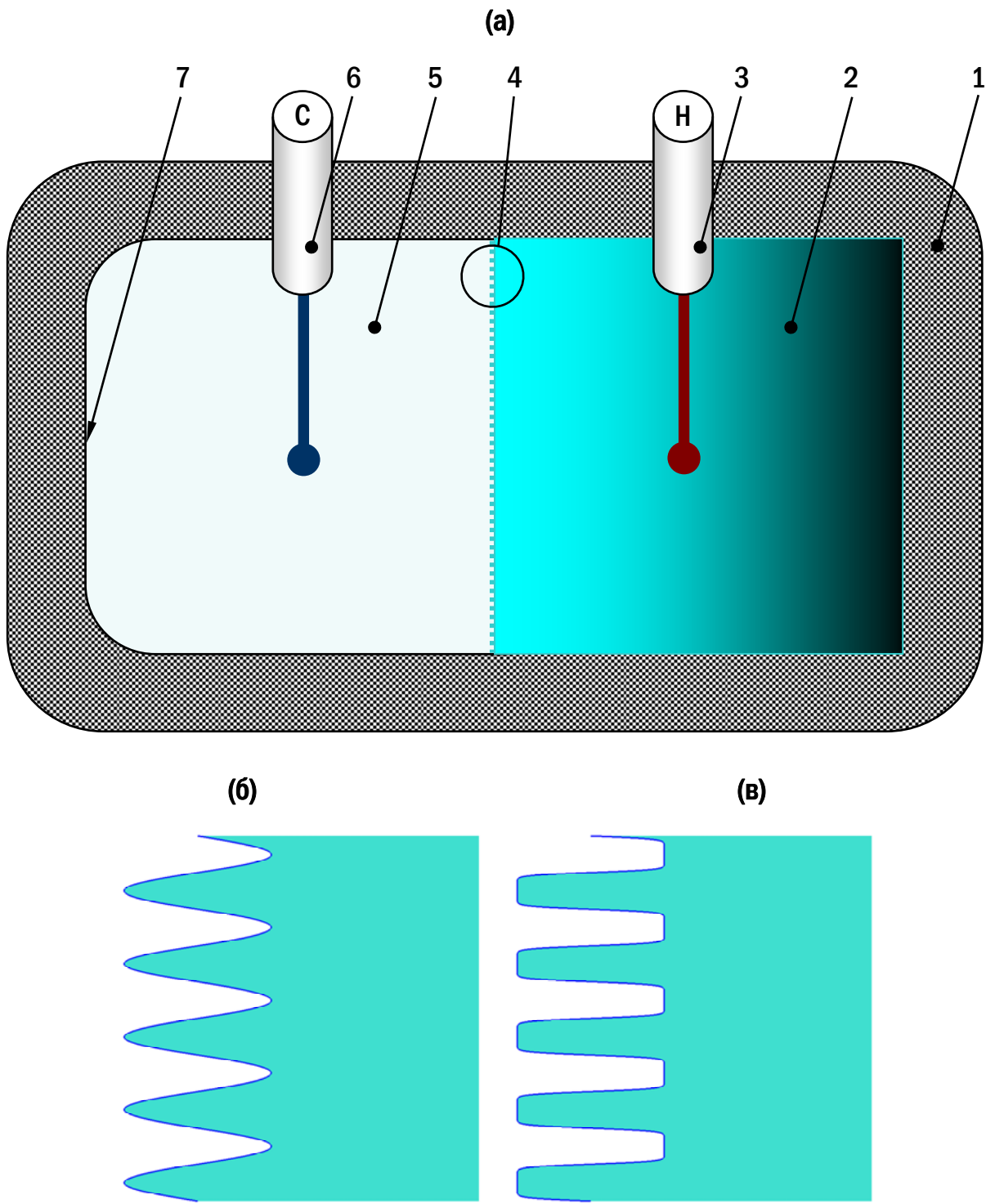

Рис. 6. Схема демонстрационной установки (а) на основе брэгговской диэлектрической решётки. Пример синусоидальной (б) и квазипрямоугольной (в) геометрии микрорельефа, для которых прогнозируется разная эффективность работы (см. оценку масштаба эффекта).

### Перечень элементов установки:

1. Термостатированный корпус установки, ограничивающий квазизамкнутую систему. Толстый слой теплоизоляции минимизирует воздействие внешних термических потоков на объекты, размещённые внутри этого корпуса.

2. Полупрозрачная дифракционная решётка с комплексной диэлектрической проницаемостью $\tilde{\varepsilon} = (n + \kappa i)^2$ и малым, но не равным нулю коэффициентом экстинкции κ.

3. Термодатчик, фиксирующий температуру решётки 2.

4. Профиль микрорельефа решётки 2. На рис. 6б и 6в изображены его синусоидальный и квазипрямоугольный варианты; см. в этом Приложении оценку масштаба эффекта.

5. Пространство внутри корпуса 1, не занятое решёткой 2. В наиболее вероятном стационарном состоянии физической системы здесь формируется излучение, плотность которого прогнозируется меньшей, чем при термодинамическом равновесии.

6. Термодатчик, фиксирующий температуру электромагнитного излучения в объёме 5.

7. Зеркальная металлическая облицовка внутренней поверхности корпуса 1, смежной с объёмом 5. Служит для минимизации воздействия на излучение внутри этого объёма каких-либо объектов, кроме решётки 2.

Работа вышеописанного устройства базируется на следующем результате имитационных экспериментов. Субволновые диэлектрические решётки с глубоким профилем микрорельефа[1], имеющие комплексный показатель преломления $\tilde{n} = n + \kappa i$ с малым, но не равным нулю коэффициентом экстинкции "κ" (показателем объёмного поглощения), обладают особыми свойствами, а именно: в ходе обмена тепловым излучением между такой решёткой и окружающим её замкнутым пространством должно установиться наиболее вероятное стационарное состояние, отличное от состояния термодинамического равновесия. Ожидаемое отличие характеризуется более низкой объёмной концентрацией фотонов в данном пространстве, чем это предполагается при исходной температуре, соответствующей изначально приготовленному равновесному состоянию. При таком развитии событий в реальной установке её термодатчики должны зафиксировать разность температур между решёткой и взаимодействующим с ней излучением.

Главную роль в механизме возникновения прогнозируемого эффекта играет конструктивная интерференция фотонов, испытывающих дифракцию Брэгга во время своего нахождения в зоне микрорельефа решётки. При этом в одинаковой степени важна возможность фотонов либо преодолеть границу диэлектрика с внешней средой и выйти наружу, либо испытать внутреннее отражение и остаться в диэлектрике: максимальная эффективность моделируемых энергоустройств, работа которых была основана на использовании данного механизма, достигалась именно при равной вероятности обоих вышеописанных исходов (брэгговский угол для первого порядка равен $\theta = 45°$).

Красными линиями на рис. 7 представлены графики плотности вероятности обнаружения в объёме 5 фотонов с заданной длиной волны в вакууме. Синими линиями обозначено исходное планковское распределение фотонного газа. Из этого материала следует, что для возникновения ожидаемого эффекта необходимо одновременное наличие микрорельефа и ненулевой экстинкции у диэлектрического оптического элемента.

---

[1] Глубокий профиль микрорельефа диэлектрической решётки фактически образует распределённый брэгговский отражатель, который, в данном случае, имеет открытую боковую стенку, совпадающую с макроповерхностью вышеупомянутой решётки.

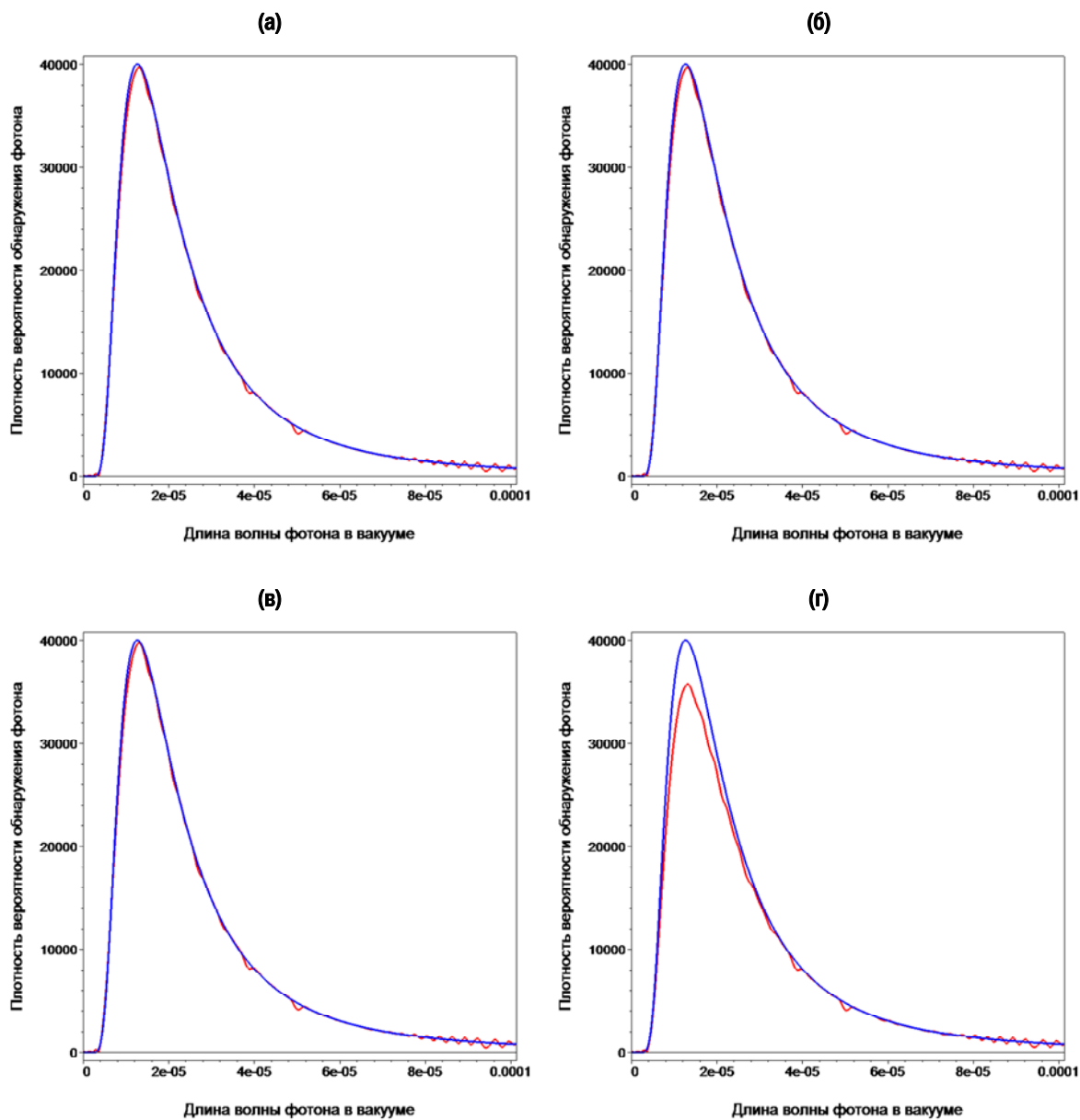

Рис. 7. Наиболее вероятные спектральные распределения фотонов в области замкнутой системы, находящейся за пределами основного оптического элемента (стационарные состояния):
 а). Микрорельеф на поверхности элемента отсутствует.   Коэффициент экстинкции к = 0.
 б). Микрорельеф на поверхности элемента присутствует. Коэффициент экстинкции к = 0.
 в). Микрорельеф на поверхности элемента отсутствует.   Коэффициент экстинкции к > 0.
 г). Микрорельеф на поверхности элемента присутствует. Коэффициент экстинкции к > 0.

Такое сочетание имеет место в системе, график плотности вероятности для которой изображён на рис. 7г. На этом графике видно, что снижение объёмной концентрации фотонов в свободной области 5 ожидается на участке исходного планковского спектра, соответствующего максимальной мощности излучения.

Рекомендуемые значения параметров диэлектрической решётки:

Материал решётки должен иметь максимально широкий спектральный диапазон внутреннего пропускания равновесного теплового планковского излучения с заданной рабочей температурой[1]. Вещественная часть комплексного коэффициента преломления $\tilde{n} = n + \kappa i$ при этом рекомендуется равной $n \approx \pi/2$. В данном случае у фотонов, находящихся в теле планарной брегговской решётки, при попытке выхода наружу имеется примерно одинаковая вероятность либо испытать полное внутреннее отражение (TIR - Total internal reflection), либо попасть в область внутренних углов падения на границе сред, когда, в отличие от варианта TIR, есть ненулевой шанс выйти из диэлектрика в свободную от оптического элемента область замкнутой системы. Для коэффициента экстинкции рекомендуется диапазон $10^{-3} < \kappa < 10^{-2}$ (возможно, придётся легировать материал диэлектрика с целью некоторого снижения его прозрачности). Вышеуказанная область допустимых значений κ, в свою очередь, соответствует диапазону вероятности $P$ того, что фотон может без поглощения пройти в диэлектрике расстояние, которое в десять раз превышает длину его волны в вакууме: $0.9875 < P < 0.9987$. Такая величина свободного пробега фотонов позволяет рассматривать динамику их прибытия из толщи решётки в область её микрорельефа аналогично тому, как это принято делать для объектов удалённой зоны (например, при дифракции по Фраунгоферу). До начала взаимодействия с внутренней поверхностью микрорельефа фотоны считаются квантовыми частицами с плоским волновым фронтом, пребывающими в чистом состоянии.

Шаг решётки $d$ должен быть в 2 раза меньше длины волны λ в материале диэлектрика, соответствующей максимуму мощности излучения для планковского спектра. При T ≈ 290°K и коэффициенте $n \approx \pi/2$ такой шаг будет равен $d = 3.183$ мкм. Полная глубина профиля $h$ микрорельефа должна превышать шаг $d$ не менее, чем в 1.5 раза.

Заметим, что для монохромного случая также достаточно оптимальное[2] соотношение $d = \lambda/2$ означает отсутствие каких-либо ненулевых дифракционных порядков рассеяния излучения как внутри диэлектрической решётки, так и снаружи её. Из этого следует, что механизм возникновения прогнозируемого эффекта здесь никак не связан с "расщеплением" фазовых траекторий фотонов на границах сред их распространения.

Для профиля микрорельефа решётки рекомендуется прямоугольная геометрическая форма. В этом случае процесс конструктивной интерференции, определяемый условием Брэгга-Вульфа, который, как предполагается, необходим для функционирования установки, реализуется гораздо более эффективно, чем при не очень глубоком ($1.5 \leq h/d \leq 2.0$) синусоидальном профиле. Электромагнитные волны, отвечающие названному условию, формируются в пространстве между взаимно параллельными отражающими плоскостями, образуемыми границами сред с различными коэффициентами преломления. В случае прямоугольного микрорельефа указанная параллельность плоскостей имеет очевидный характер. Для синусоидального профиля можно говорить о некотором приближённом подобии такой параллельности, если глубина $h$ микрорелье-фа решётки относительно её шага *d* достаточно велика: $h/d \gg 1$. Замена этого профиля на прямоугольный должна увеличить энергоотдачу описываемого здесь устройства.

---

[1] Весьма эффективным (но не самым технологичным из-за растворимости в воде) материалом для такой решётки может быть бромид калия KBr. Он имеет близкий к оптимальному коэффициент преломления n ≈ 1.527 и аномально большой размер окна внутреннего пропускания: 98.82% для теплового излучения с температурой 290°K.

[2] Для монохромного случая строго оптимальным является соотношение $d/\lambda = \sqrt{2}/2$.

Оценим масштаб прогнозируемого эффекта на конкретном примере: для варианта решётки из легированного (к = 0.005) бромида калия KBr, имеющей шаг синусоидального[1] микрорельефа *d* = 3.46 мкм с полной глубиной *h* = 10.48 мкм, уменьшение объёмной концентрации фотонов в свободной от решётки области составит ≈ 3.58%. При исходной температуре термодинамического равновесия 290°K это соответствует разнице температур между решёткой и излучением в свободной области ~ 2.63°K.

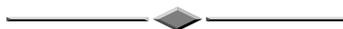

---

[1] Замена синусоидального профиля микрорельефа на прямоугольный (см. рис. 6б и 6в, геометрические пропорции микрорельефа на рисунках соответствуют расчётным данным) позволяет повысить предполагаемую энергоотдачу устройства в ≈ 3.46 раза.

## Список использованной литературы

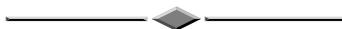